# Moiré-free ultrathin iron oxide film: FeO(111) on Ag(111)

M. Lewandowski,[a,*] T. Pabisiak,[b,**] N. Michalak,[c] Z. Miłosz,[a] V. Babačić,[a] Y. Wang,[a] M. Hermanowicz,[d] K. Palotás,[e] S. Jurga,[a] A. Kiejna[b] and R. Wiesendanger[f]

a. NanoBioMedical Centre, Adam Mickiewicz University, Umultowska 85, 61-614 Poznań, Poland
b. Institute of Experimental Physics, University of Wrocław, pl. M. Borna 9, 50-204 Wrocław, Poland
c. Institute of Molecular Physics, Polish Academy of Sciences, M. Smoluchowskiego 17, 60-179 Poznań, Poland
d. Institute of Physics, Poznan University of Technology, Piotrowo 3, 60-965 Poznań, Poland
e. MTA-SZTE Reaction Kinetics and Surface Chemistry Research Group, University of Szeged, 6720 Szeged, Hungary
f. Department of Physics, University of Hamburg, Jungiusstr. 11a, 20355 Hamburg, Germany

*,** Corresponding Authors:
lewandowski@amu.edu.pl (experiment), tomasz.pabisiak@uwr.edu.pl (theory)

**Electronic Supplementary Information (ESI) available:** Literature data concerning preparation of iron oxide films on silver single crystal supports, detailed description of experimental apparatus, procedures and theoretical calculations, detailed x-ray photoelectron spectroscopy data, geometrical details of the calculated and experimentally observed systems, calculated Bader charges, magnetic moments and work function values, additional figures. See DOI: 10.1039/x0xx00000x

**Abstract**

Ultrathin iron oxide films epitaxially grown on the (111)- and (0001)-oriented metal single crystal supports exhibit unique electronic, catalytic and magnetic properties not observed for the corresponding bulk oxides. These properties originate mainly from the presence of Moiré superstructures which, in turn, disqualify ultrathin films as model systems imitating bulk materials. We present a route for the preparation of a close-packed Moiré-free ultrathin iron oxide film, namely FeO(111) on Ag(111). Experimental scanning tunneling microscopy (STM), low energy electron diffraction (LEED) and x-ray photoelectron spectroscopy (XPS) results confirm perfect structural order in the film. Density functional theory (DFT)-based calculations suggest full relaxation of the oxide layer that adopts the atomic lattice of the crystalline support and exhibits properties similar to those of a free-standing FeO. The results open new pathways for model-type studies of electronic, catalytic and magnetic properties of fully-relaxed iron oxide films and related systems.

**Introduction**

Ultrathin films are believed to constitute a new class of 2D materials with properties governed by low-dimensionality and interaction with the support and, therefore, different from those of the corresponding bulk materials.[1] Historically, ultrathin insulating films grown on conducting supports were meant to be used as model systems that would allow the use of surface science tools for the studies of materials that are insulators in the bulk form.[2] However, the preparation of epitaxial ultrathin films with properties that are not governed by lattice mismatch-induced superstructures was found to be challenging.

Monolayer iron oxide (FeO) films can be grown in the close-packed [111] direction on various metal single crystal supports.[3–7] Until now, all of these films exhibited lattice mismatch-induced Moiré superstructures that were determining their properties and making them not representative as model systems imitating bulk oxide material. Silver has always been considered a promising substrate for iron oxides growth, as it has the same crystal symmetry as FeO, only ~5% lattice mismatch, is considered a weakly-interacting substrate and is relatively resistant to oxidation. In addition, silver was successfully used as a substrate for the growth of another reconstruction-free ultrathin oxide film – MgO(001).[8]

There were several attempts to grow iron oxide films on silver single crystal supports.[9–20] A table that summarizes these attempts is presented in the Electronic Supplementary Information, section 1 (ESI1). However, most of the studies concerning ultrathin FeO(111) films on Ag(111) were contradictory and lacked detailed structural characterization by atomic-resolution techniques. Some works suggested a (1×1) growth of FeO[9] (without providing a convincing proof), other showed the formation of Moiré structures (see e.g. Refs. 18 and 19). The authors often denoted these films as "FeO$_x$", not "FeO", thus indicating their ill-defined character.

In this work, we present a route for the preparation of a Moiré-free ultrathin iron oxide film – the FeO(111) on Ag(111). Such film has never been observed on any close-packed metal surface. We used atomic-resolution scanning tunneling microscopy (STM), low energy electron diffraction (LEED) and monochromated x-ray photoelectron spectroscopy (XPS) in combination with density functional theory (DFT)-based calculations, to analyze the structure of the film. The experimental results confirm perfect structural order in the FeO layer, while DFT calculations suggest full relaxation of the oxide film that adopts the atomic lattice of the crystalline support. The calculations also show that the film preserves the properties of free-standing FeO, thus making it representative as model system imitating bulk oxide.

**Experimental and theoretical methods**

The experiments were performed in an ultra-high vacuum (UHV) chamber (pressure ≤ $10^{-10}$ mbar) equipped with standard sample cleaning (Ar$^+$ sputter gun, e-beam heating stage, evaporators, oxygen line) and characterization (STM, LEED, XPS) surface science tools. Ultrathin FeO(111) films were grown by Fe deposition from a 2 mm rod onto a Ag(111) single crystal support held at 500-600 K and subsequent oxidation in 1×10$^{-6}$ mbar O$_2$ at 700 K for several minutes. The cleanliness of the substrate, as well as the structure of Fe and FeO deposits, were characterized by STM, LEED and XPS. All STM images were obtained using electrochemically etched W tips. The dI/dz curves were recorded using a lock-in technique.

The theoretical calculations were based on DFT, as implemented in the Vienna ab initio simulation package (VASP).[21,22] The electron-ion core interactions were represented by the projector-augmented wave (PAW) potentials.[23,24] The exchange-correlation energy was treated at the generalized gradient approximation (GGA) level, using the Perdew-Burke-Ernzerhof (PBE) functional.[25] To account for the strong correlation of 3d electrons localized on the Fe ions, the Hubbard correction term U has been applied (GGA+U) within the rotationally invariant approach of Dudarev et al.[26] by using an effective parameter $U_{eff}$ = 3.0 eV, which is the difference between the Coulomb, U, and exchange, J, parameters. The convergence threshold for the total energy of the studied systems was set to $10^{-6}$ eV. A silver (111) substrate was modelled with an asymmetric slab consisting of four atomic layers of Ag, separated from its periodic images by a vacuum region of 18 Å. The positions of atoms in the bottom two Ag layers were frozen and the atomic positions of the remaining atoms were optimized. An FeO(111) monolayer was adsorbed on one side of the Ag(111) slab. For the initial structural modelling, the calculated in-plane lattice constants of 2.938 Å for Ag(111) (PBE) and 3.032 Å for FeO(111) (PBE+U) were used. The STM image was simulated using the Tersoff-Hamann method.[27]

The detailed description of the experimental apparatus, procedures and theoretical calculations can be found in ESI2.

**Results and discussion**

Figure 1a presents a large-scale STM topography image and a LEED pattern (inset) of a monolayer (ML) FeO(111) film grown on Ag(111) by ~1 ML iron deposition at 550 K and subsequent oxidation in $1 \times 10^{-6}$ mbar $O_2$ at 700 K for several minutes. The image reveals smooth surface with no visible signs of lattice mismatch-induced superstructures. The height of the film, determined from several height profiles taken across deepest holes that could be found on the acquired images, is ~3.3 Å. The real height of the film can slightly differ, as the measured value may depend on the local electronic structure and the bias voltage used, which is particularly important when measuring height differences between oxide and metal surfaces. The height difference between neighboring terraces is ~2.35 Å, corresponding to the height of a monoatomic step on Ag(111).[28] These values are indicative for the presence of an iron oxide layer of uniform height, dominantly covering the atomic terraces of the silver support. On some samples we also observed the formation of second FeO layer islands (see Figure S1a in ESI). The height of such islands is smaller and equals ~2.5 Å.

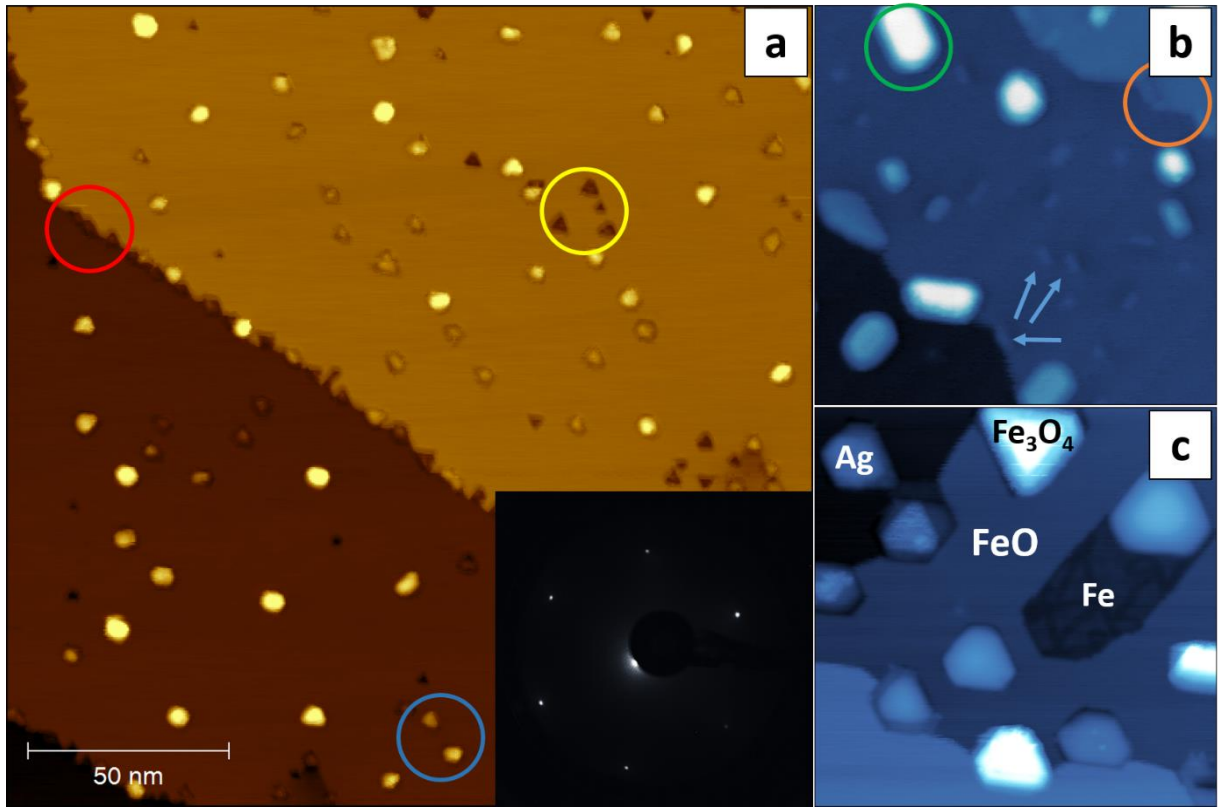

**Fig. 1** Large-scale topographic STM image (200×200 nm$^2$) of a monolayer FeO(111) film on Ag(111) (V = +1.0 V, I = 0.7 nA) (a); Inset: LEED pattern (60 eV) exhibiting a (1x1) structure; (b) and (c) present topographic STM images (50×50 nm$^2$) of 550 K-grown iron structures and different structures observed after oxidation, respectively (V = -3.0 V (a) and +0.7 V (b), I = 0.4 nA) (see text for details).

Following other authors,[4] we tentatively assign the FeO film to be O-terminated. The observed cleanliness of the surface is indirectly supporting this assumption, as O-terminated FeO(111) surfaces are usually chemically inert.[29] On some samples, triangular and hexagonal defects (marked with a yellow circle in Figure 1a), with edges running along $\sqrt{3}$ crystallographic directions with respect to the directions of the Ag(111) support, were observed. As the depth of most of these defects is smaller than the determined film height, we tentatively assign them to missing oxygen atoms resulting in exposed iron (for atomic resolution image of such defect please see Figure S1b). We also observed thicker particles that were sticking out from the film (Figure 1a, blue circle). The nature of these particles is not known, however, we believe they are silver.

The edges of the film have a sawtooth-like shape (Figure 1a, red circle) and run along similar crystallographic directions as the defects' edges. Notably, the orientation of the edges indicate that the <110> direction, characteristic for step edges of (111)-oriented surfaces,[30] is not preferred for this film, while the $\sqrt{3}$ directions are. It has to be mentioned that in contrast to most other works on the growth of iron oxide films on Ag(111) (see e.g. Refs. 9, 18 and 19), our preparation procedure did not involve room-temperature growth of iron, which results in ill-defined iron clusters located mainly at the Ag(111) step edges,[31] but high-temperature growth of bcc-Fe crystals at the terrace sites of silver (with only small fraction of iron particles growing at the silver step edges).[32] We found that high-temperature iron deposition is a crucial step for well-ordered FeO(111) film preparation on this particular support. Deposition of sub-monolayer amounts of iron onto a heated silver substrate results initially in the formation of small, elongated, rhomboidal iron inclusions at the terrace sites of silver (Figure 1b, marked with arrows). The fraction of iron particles growing at the silver step edges was found to additionally modify the steps' direction (Figure 1b, orange circle). The observed inclusions have a height of ~(0.6-1.2) Å (approx. half of the height expected for monolayer bcc-Fe, i.e. 2 Å)[30] and top facets tilted with respect to the substrates' plane – thus suggesting that they are embedded in the surface. This may appear surprising, taking into account the fact that iron generally does not alloy with silver. However, it was shown that diffusion of iron into silver bulk is possible at high temperatures,[33] therefore, we assume that at temperatures > 500 K iron could be incorporated into the top silver layer. At higher coverages, these inclusions most probably act as nucleation centers for larger iron bcc nanocrystals which grow in the <110> direction at the terrace sites of silver (Figure 1b, green circle) and, following oxidation, facilitate the growth of continuous iron oxide films thanks to the uniform distribution of iron at the surface. Interestingly, similar metal incorporation at the oxide/support interface was also observed by other authors during the growth of MgO(001) films on Ag(001).[8,34] We speculate, therefore, that the formation of inclusions is a critical intermediate step for well-ordered oxide film growth on silver substrates. It also has to be mentioned that in our experiments, room-temperature deposition of iron resulted in silver step erosion and vacancy island formation (similarly to the reported case of copper[35]). Significant mass transport could facilitate the formation of the observed silver inclusions embedded in the FeO(111) films. In addition to silver inclusions, we also rarely observed small islands exhibiting hexagonal shape and large in-plane atomic periodicity (see Figures 1c and S1c). These islands were tentatively assigned to the initial stages of $Fe_3O_4$(111) growth. It is expected that the fraction of defects, embedded particles and $Fe_3O_4$ islands, could be adjusted by fine-tuning the preparation conditions.

Even though STM results indicated that the surface is dominantly covered with an iron oxide film, it still exhibited a (1×1) LEED pattern similar to that of Ag(111), with no signatures of lattice mismatch-induced superstructures (as determined from diffraction patterns recorded at various energies ranging from 30 to 255 eV). The presence of a (1×1) LEED pattern was also reported by other authors following room temperature Fe deposition onto Ag(111) and post-oxidation (see e.g. Ref. 9), however, the spots they observed were rather diffuse, while our LEED spots were as sharp as one could expect for a clean, reconstruction-free (111) surface. In addition, several spot profiles taken over different diffraction spots on the acquired LEED patterns confirmed the absence of beam-splitting in our case. The presence of significant amounts of other iron oxide phases, i.e. $Fe_3O_4$(111) or $Fe_2O_3$(0001), was excluded, as they would lead to (2×2) or ($\sqrt{3} \times \sqrt{3}$)R30° LEED patterns, respectively.[4,14,16]

Figure 2a presents XPS Fe 2p data obtained for the FeO(111) film shown in Figure 1a. The analysis of the spectra was not trivial, as the Fe 2p region overlaps with the Ag 3s signal (presented for clean Ag(111) in Figure 2b). In order to obtain reliable results, we fitted the original data (Figure 2a) and, in addition, the data after the subtraction of the Ag 3s signal from Figure 2b (the spectrum after subtraction is shown in Figure 2c). As both fittings (Figures 2a and 2c) gave similar results, we only discuss here the features of the Fe 2p spectrum after Ag 3s subtraction, while detailed information on both fittings is provided in ESI3.

The Fe 2p signal could be fitted with six components. The 710.3 eV and 723.9 eV peaks, chemically shifted by approx. (3.5-4.5) eV to higher binding energies with respect to metallic iron (706.8 eV and 719.8 eV),[36] result from $Fe^{2+}$ ions in iron oxide. The 717.5 eV and 731.2 eV components correspond to characteristic $Fe^{2+}$ satellite peaks.[37] The peaks centered at around 713.2 eV and 727.4 eV were assigned to result from multiplet splitting.[38,39] Metallic iron or iron in the $Fe^{3+}$ oxidation state,[36] if any, was beyond the detection limit of our XPS system. The lack of metallic iron confirms that that the particles embedded in the oxide film are silver. The recorded oxygen O 1s signal consists of a single peak centered around 529.7 eV[40] and is presented in Figure S2. The Fe to O ratio, determined from a survey spectrum (not shown) and by taking the elemental photoionization cross-sections[41] into account, was approx. 1:1 – as expected for the FeO phase with perfect stoichiometry. No other elements, like e.g. carbon, sulfur or other contaminants, could be detected. Also, no significant changes in the Ag 3d peak shapes were observed, which indicated weak oxide-support interaction. All these confirm the presence of an FeO film weakly bound to the Ag substrate.

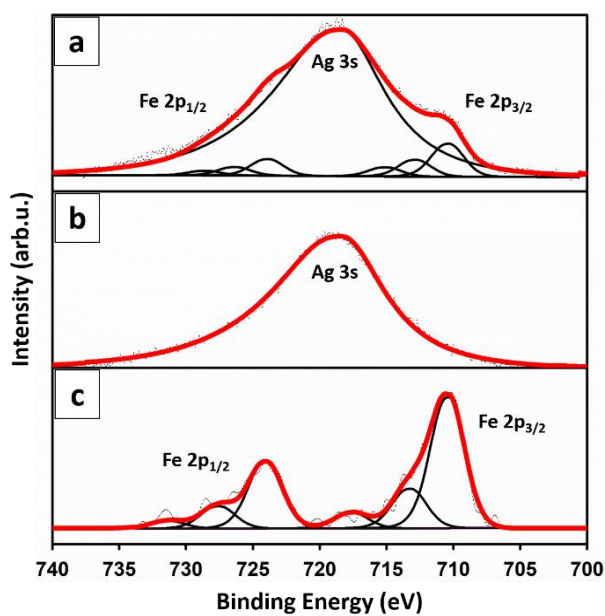

**Fig. 2** XPS Fe 2p spectra of FeO(111) on Ag(111), before (a) and after (c) subtraction of the Ag 3s signal obtained for clean Ag(111) (b).

In our periodic total energy calculations the FeO(111) monolayer on a Ag(111) surface was modelled using a 1×1 surface unit cell, thus adopting the lattice constant of the support. The results were obtained for the calculated (PBE) Ag bulk lattice constant of 4.155 Å and FeO lattice constant of 4.288 Å (PBE+U), which result in a 3.1% misfit of (111) surface lattice constants. It has to be mentioned that the experimental lattice parameters of Ag (4.086 Å) and

FeO (4.32 Å) result in a (111) surface lattice mismatch of 5.7%. Due to this, a similar set of calculations, including structural optimizations of considered mono- and bilayer FeO(111) systems, was performed using the experimental lattice constants of Ag and FeO. Importantly, the calculated geometries, presented in ESI4, do not show substantial differences with respect to those obtained with theoretically determined lattice constants.

Several configurations of Fe- and O-terminated FeO(111)/Ag(111) systems were examined, by considering both Fe and O atoms placed in fcc or hcp hollows, as well as on top of Ag atoms. The most stable Fe-termination is the one with Fe atoms occupying fcc positions and O atoms in hcp hollow sites (with an O-Ag distance of 2.45 Å and an Fe-O distance of 0.9 Å). Other Fe-terminations are at least 10 meV less favored. The most stable O-termination is again the one with Fe and O atoms sitting in the threefold coordinated fcc and hcp hollow sites of the Ag(111) surface, respectively (with an Fe-Ag distance of 2.35 Å and an O-Fe distance of 0.82 Å). This configuration (presented in Figure 3 (left)) is by 0.79 eV energetically more favored than the Fe-terminated one – in agreement with a tentative experimental assumption on the oxide's surface termination. It has to be mentioned that for the 1×1 supercell calculations, this configuration is nearly degenerated in energy with the one with Fe in hcp and O in fcc sites (which is less stable by only 2 meV). However, the calculations performed using a 2×2 supercell confirmed the same preference of adsorption sites with 6 meV difference, and those performed using the experimental lattice constant of Ag (4.086 Å) showed that the two configurations differ by 15 meV. Another O-termination, with O atoms in on-top positions and Fe atoms in either fcc or hcp sites, is by 89 meV less favored. Due to this, further calculations were performed for the O-terminated structure with Fe in the threefold coordinated fcc sites and O in the hcp hollow sites.

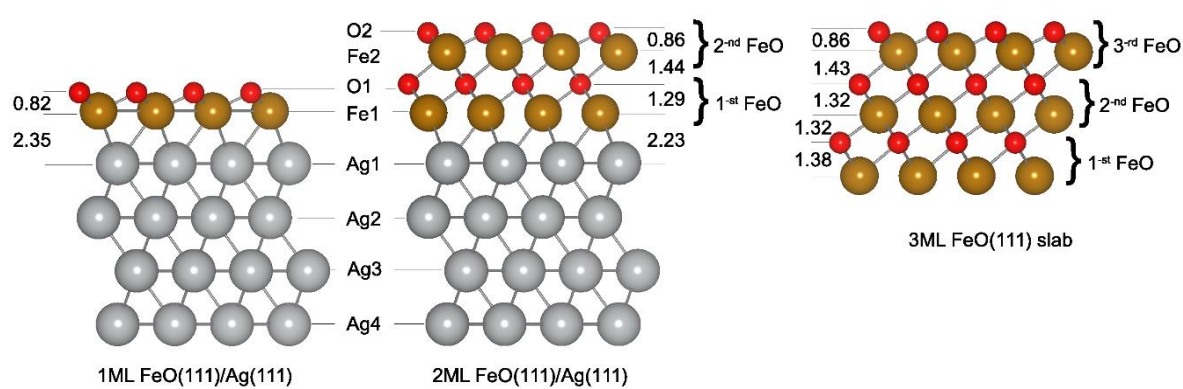

**Fig. 3** Side views of O-terminated monolayer (left) and bilayer (middle) FeO(111) films on Ag(111), as well as 3-MLs-thick FeO slab (right), determined from PBE+U calculations using a 1×1 surface unit cell with fixed 2.938 Å FeO(111) and Ag(111) in-plane lattice constants (FeO(111) film adopting the geometry of the Ag(111) substrate). All distances are given in Å.

The calculated total height of the O-terminated monolayer FeO(111) film was found to be 3.17 Å. This value is close to the experimentally observed one (~3.3 Å). A strongly reduced Fe-O interlayer spacing in the FeO(111)/Ag(111) system with respect to bulk FeO (1.25 Å) is of similar magnitude to that reported for FeO(111)/Pt(111)[42,43] and results from a complex stabilization mechanism, i.e. interplay between structural relaxation, polarity compensation, charge transfer and magnetic ordering[43–48] which in the case of FeO(111)/Ag(111), unlike in the case of FeO(111)/Pt(111), does not lead to the development of a Moiré-type coincidence structure. An inspection of the calculated Bader charges (see ESI5 and Figures S3) on the interface atoms of the FeO(111) monolayer deposited on a Ag(111) substrate shows that adsorption of an FeO film results in the appearance of negative charges on silver (0.17 |e|) and iron (0.04 |e|) atoms, gained at the expense of oxygen (0.24 |e|). This charge transfer may be important with respect to film's stabilization. Another stabilizing factor may be the magnetic superstructure. It is worth noting that the applied 1×1 (or a larger 2×2) surface unit cell gives a ferromagnetic (FM) ground state and does not allow reproducing an antiferromagnetic (AFM) superstructure that results from magnetic frustration of antiferromagnetic FeO on the three-fold symmetric oxide layer.[44] Such a magnetic superstructure is observed indeed if a 3×3 or a ($\sqrt{3} \times \sqrt{3}$)R30° surface unit cell is applied, where the number of Fe atoms in a layer is a multiple of three with a 1:2 ratio of Fe atoms with opposite magnetic moments. Such an AFM structure is by 0.26 eV per FeO unit more stable than the FM one. However, the AFM alignment of magnetic moments on Fe atoms makes the iron layer distinctly rumpled, with Fe atoms of one direction of magnetization being by around 0.5 Å more distant from the Ag(111) surface than those with opposite magnetization (Figure S4). The iron oxide layer observed experimentally does not exhibit such rumpling (not at the large-scale – as shown above, nor at the atomic scale – as will be shown below), at least at temperatures exceeding the Néel temperature of FeO (~198 K[49]). Therefore, the magnetic superstructure resulting from magnetic frustration can be considered as not appropriate to describe the FeO films observed in our room-temperature experiments.

The bilayer FeO(111) film was modelled by placing a second monolayer of FeO(111) on top of an FeO(111)/Ag(111) slab, originally in a flat configuration, with Fe over surface Ag atoms and O in fcc positions over Fe atoms in the first FeO layer (Figure 3 (middle)). Upon structural optimization, the lateral positions of Fe and O atoms remained unchanged and the atoms relaxed only vertically. Both FM and AFM stackings of the ferromagnetically ordered adjacent FeO layers were considered. The AFM configuration appeared to be by 63 meV more stable and was adopted in further calculations. The height of the first (interface) monolayer was increased to 3.52 Å due to a substantial increase of the O-Fe distance (to 1.29 Å). The thickness of the second (top) monolayer was 2.30 Å (experimentally determined value: ~2.5 Å), with an O-Fe distance of 0.86 Å. The distance between the iron oxide layers was 1.44 Å and the distance between the FeO bilayer and the silver support was 2.23 Å. The relatively large Fe-Ag distances suggest weak interaction of the oxide film with the silver support. Importantly, the calculated first- and second-monolayer heights were found to be in fair agreement with the values observed experimentally. It also has to be mentioned that the results obtained for mono- and bi-layer FeO(111) films did not change with a change of the supercell from 1×1 to 2×2.

In order to check the influence of the Ag(111) support on the structure of FeO(111)

films, we also simulated a free-standing 3-monolayer-thick (6 atomic layers) FeO(111) slab and compared its parameters with those of a Ag(111)-supported film (see Figure 3 (right) and ESI4). An antiferromagnetic configuration of alternating layers was found to be by 2.01 eV more favored than the ferromagnetic one, so further description is given for the AFM phase only. The slab is asymmetric due to differently terminated sides which makes the interlayer separations at the O- and Fe-terminated sides, respectively, shrink or expand. Notably, the distances between the inner Fe and O planes (1.32-1.43 Å) were found to be close to the Fe-O spacing in a bilayer FeO(111) on Ag(111) (1.29-1.44 Å) and in bulk FeO (1.25 Å). It can be thus concluded that the second FeO(111) layer on Ag(111) has a "free-standing FeO" character, which also explains the absence of polarity-induced reconstructions (which are present e.g. in the bilayer FeO(111)/Pt(111) system[50,51]). The similarity of Ag(111)-supported bilayer FeO(111) films and a free-standing 3-MLs-thick FeO(111) slab was also reflected in the calculated Bader charges and magnetic moments (ESI5). The magnitude of both electron charges and magnetic moments on the atoms of the top surface layer of a bilayer film and a free-standing slab is mainly determined by the presence of the surface. In both structures, the O and Fe atoms of the outermost layer lose 0.25 and 0.41-0.43 electrons, respectively. The electron charge gain on Fe1 atoms of the bilayer (0.51 |e|) and Ag1 atoms of the substrate (0.19 |e|) again indicate the weak character of the FeO-Ag bonding. The magnetic moments on atoms of the outermost FeO layer of the bilayer and a free-standing 3-MLs-thick slab are nearly the same which means that they are determined by the presence of both the surface and the underlying FeO monolayer. In the surface FeO monolayer, the magnetic moments on oxygen atoms (0.39 µB) are induced and the moments on Fe atoms are enhanced. The presence of an underlying FeO layer leads to a further increase of moments on surface Fe and O atoms to the same magnitude, independently of the presence of the silver support.

  The partial densities of electronic states (PDOS) calculated for the FeO(111) monolayer on Ag(111), the second FeO layer from a bilayer FeO(111) film on Ag(111) and the top FeO layer from the 3-MLs-thick FeO slab, are presented in Figures 4(a-c). A similarity between the electronic structure of all three systems can be seen, especially when comparing the second FeO layer on Ag(111) to a free-standing FeO slab. It suggests that the density of states is mainly determined by the presence of the surface and not by the interaction with the Ag(111) support. However, the presence of the underlying FeO layer is important for the opening of an energy gap of 0.7-0.9 eV in the topmost FeO layer (in the majority spin states; see Figures 4b and 4c). There is also a great asymmetry in the majority and minority spin states, in particular in those belonging to Fe. The former are almost completely occupied, whereas the latter are mostly empty. The Fe-DOS is dominated by the Fe 3d-states, while the O-DOS by the O 2p-states.

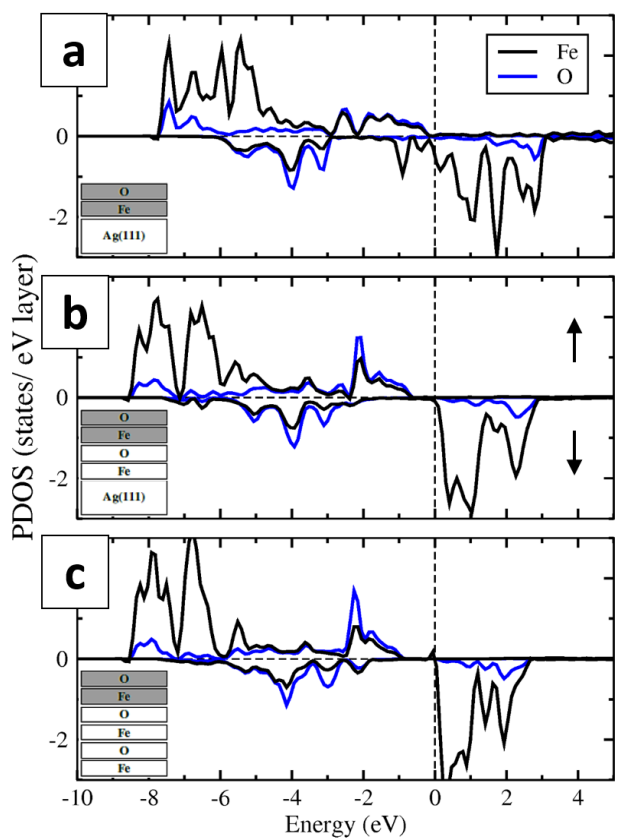

**Fig. 4** PDOS of an FeO(111) monolayer on Ag(111) (a), second FeO(111) layer from a bilayer FeO(111) film on Ag(111) (b), and the top FeO layer from a free-standing 3-MLs-thick FeO(111) slab (c), resulting from PBE+U calculations.

Atomic-resolution STM images of FeO(111)/Ag(111) are shown in Figures 5a and 5b. The structure consists of a lattice of small, sharp protrusions (marked with black circles) superimposed on a similar lattice of bigger and more diffuse appearing species (yellow circles). A similarly spaced "lattice" of holes (blue circles) can also be observed. The atomic periods within the respective lattices, determined from ~100 line profiles, were found to be (2.94 ± 0.08) Å (surface lattice constant of Ag(111): 2.89 Å). The standard deviation results from thermal drift and STM image distortion caused by room-temperature imaging at high scanning speed. It has to be mentioned that $Fe_3O_4$(111) or $Fe_2O_3$(0001) surfaces would show atomic periodicities of 6 Å and 5 Å, respectively.[4] Notably, the experimentally observed atomic structure of FeO was locally defect-free, without point defects or other imperfections, and exhibited no in-plane rumpling. Figure 5c presents a simulated STM image of FeO(111)/Ag(111). A perfect agreement between experiment and theory is achieved, where the smaller protrusions correspond to oxygen atoms, the bigger to iron atoms and the holes represent the positions of silver atoms.

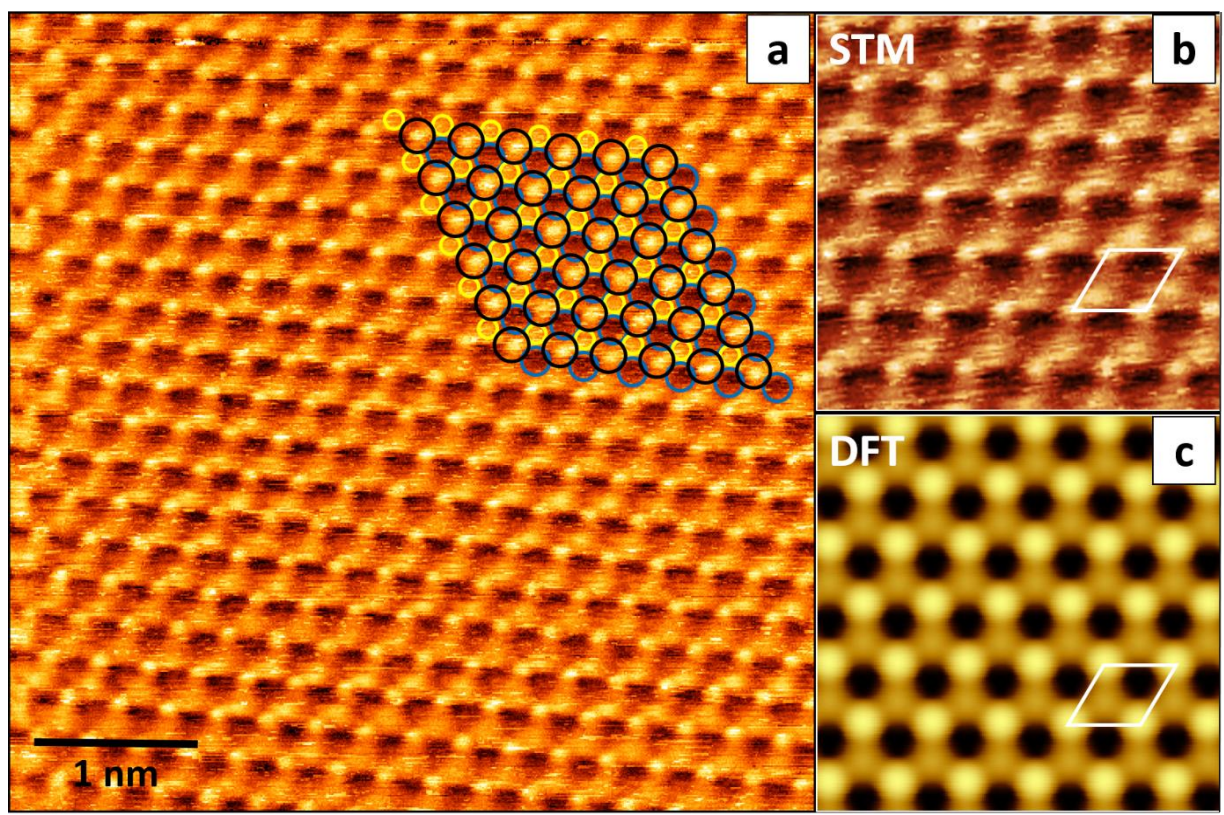

**Fig. 5** Atomic-resolution STM image (5×5 nm$^2$) of FeO(111) on Ag(111) (V = +0.1 V, I = 0.4 nA) (a). Contributions from two atomic sublattices were visible on topography and current images, however, with different intensities. Due to this, a sum of the two images is presented (for separate topography and current images see Figure S5). The positions of iron (yellow), oxygen (black) and silver (blue) atoms are marked with circles. Differences in the circles' sizes correspond to differences in ionic radii of the respective elements. (b) presents a zoom of (a) (1.73×1.73 nm$^2$). (c) shows a simulated STM image based on DFT calculations and the Tersoff-Hamann method.[27] The FeO(111)-(1×1) unit cell is marked in (b) and (c) with a white rhombus.

The calculated work function values of a monolayer FeO(111) film on Ag(111) range from 5.87 eV (AFM configuration, ($\sqrt{3} \times \sqrt{3}$)R30° unit cell) to 6.56 eV (FM configuration, 1×1 unit cell) (see ESI6 for values obtained for different configurations of magnetic arrangements and computational unit cells). As these values are much higher than the work function of clean Ag(111) (4.49 eV (calculated), 4.35-4.74 eV (experimental[52–54])), local work function measurements were performed to directly confirm that the atomically-flat regions seen on STM images in between the inclusions are indeed FeO(111) and not clean Ag(111). For this, point scanning tunneling spectroscopy (STS) dI/dz data were collected on the regions assigned to FeO(111) and compared with the data obtained for clean Ag(111), as well as other single-crystalline surface with much higher work function value: Pt(111) (5.93-6.1 eV[55,56]). The I(z) spectroscopy reassembles the exponential change of the tunneling current with changing tip-sample distance.[57] The dependence can be measured even more precisely by recording the first derivative of the signal, i.e. dI/dz, using lock-in technique.[58,59] As the slope of the dI/dz curve is proportional to the work function of the sample, the technique allows distinguishing surfaces with different work function values (assuming constant work function of the tip during all measurements). The method was successfully applied by other authors to study ultrathin iron oxide films and determine local work function variations within the Moiré superstructure of FeO(111)/Pt(111).[45]

The recorded dI/dz curves are shown in Figure S6, while Figure 6 presents excerpts from the curves plotted in a logarithmic scale. The slope of the curve obtained for FeO(111) was much higher than that of clean Ag(111) and lower than that of Pt(111). The measurements were repeated several times for different sample preparations and the trend was always the same. Based on this, it may be concluded that the flat regions seen on STM images are indeed covered with FeO(111) and that the oxide's work function value lies between 4.49 eV and 6.1 eV.

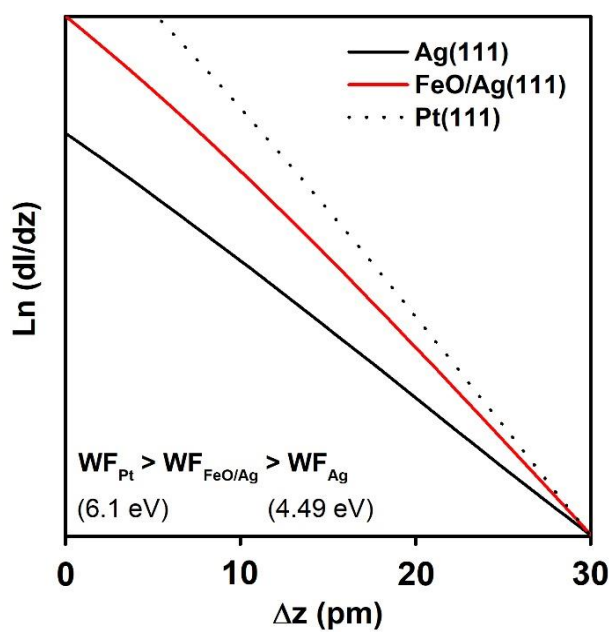

**Fig. 6** dI/dz curves obtained for FeO(111)/Ag(111), clean Ag(111) and Pt(111) (logarithmic plot).

On the way to obtain a recipe for a Moiré-free epitaxial FeO(111) film on Ag(111), we also reproduced several preparation procedures used by other authors. We performed experiments which involved iron deposition at room temperature and subsequent oxidation or deposition of iron in an oxygen ambient. Examples of the resulting structures are presented in Figure S7. However, despite numerous efforts, these approaches did not lead to the formation of a reconstruction-free iron oxide wetting layer, but to islands exhibiting Moiré superstructure or ill-defined $FeO_x$ structures.

**Conclusions**

We presented a route for the preparation of a Moiré-free ultrathin FeO(111) film on Ag(111). The procedure includes iron deposition onto a heated substrate, which assures a uniform distribution of deposited material on the surface and epitaxial growth of bcc-(110) iron crystallites, and subsequent oxidation. As prepared iron oxide films exhibit perfect structural order, as shown by STM, LEED and XPS. DFT-based calculations indicate adaptation of the oxide layer to the silver support. In addition, the calculated structural parameters and PDOS of the second FeO(111) layer on Ag(111) are similar to those of a top FeO layer in a free-standing FeO slab and in FeO bulk. This indicates that the FeO(111) film on Ag(111), besides being Moiré-free, has more bulk-like properties than FeO films grown on other supports and, thus, can be representative as a model system that imitates bulk FeO. The results contribute to the general knowledge on epitaxial oxide films and open new pathways for model-type studies of electronic, catalytic and magnetic properties of fully-relaxed iron oxide films and related systems.

**Acknowledgments**

This work was financially supported by the National Science Centre of Poland (SONATA programme, 2013-2017, grant No. 2012/05/D/ST3/02855, MAGNETON project) and the Foundation for Polish Science (First TEAM programme, 2017-2020, grant No. First TEAM/2016-2/14 ("Multifunctional ultrathin Fe(x)O(y), Fe(x)S(y) and Fe(x)N(y) films with unique electronic, catalytic and magnetic properties" project co-financed by the European Union under the European Regional Development Fund)). T.P and A.K. acknowledge computer time granted by the ICM of the Warsaw University (Project G44-23).[60]

**Notes and references**

Author contributions: M.L. performed STM, LEED and XPS measurements, data analysis and wrote the manuscript. N.M. and Z.M. assisted in STM, LEED, XPS measurements and data analysis. V.B. and Y.W. performed dI/dz measurements supervised by M.L. T.P. performed all numerical calculations (with contribution of M.H. in the early stage of the work) and analyzed the results together with A.K., who wrote the theoretical part of the manuscript. K.P. contributed to the discussion of simulated STM images. S.J. and R.W., in addition to all other authors, contributed to the discussion, where R.W. authored the original idea of Moiré-free ultrathin iron oxide film preparation.

# Electronic Supplementary Information (ESI)

# Moiré-free ultrathin iron oxide film: FeO(111) on Ag(111)

M. Lewandowski,[a*] T. Pabisiak,[b**] N. Michalak,[c] Z. Miłosz,[a] V. Babačić,[a] Y. Wang,[a] M. Hermanowicz,[d] K. Palotás,[e] S. Jurga,[a] A. Kiejna[b] and R. Wiesendanger[f]

*a. NanoBioMedical Centre, Adam Mickiewicz University, Umultowska 85, 61-614 Poznań, Poland*
*b. Institute of Experimental Physics, University of Wrocław, pl. M. Borna 9, 50-204 Wrocław, Poland*
*c. Institute of Molecular Physics, Polish Academy of Sciences, M. Smoluchowskiego 17, 60-179 Poznań, Poland*
*d. Institute of Physics, Poznan University of Technology, Piotrowo 3, 60-965 Poznań, Poland*
*e. MTA-SZTE Reaction Kinetics and Surface Chemistry Research Group, University of Szeged, 6720 Szeged, Hungary*
*f. Department of Physics, University of Hamburg, Jungiusstr. 11a, 20355 Hamburg, Germany*

\*,\*\* Corresponding Authors:
lewandowski@amu.edu.pl (experiment), tomasz.pabisiak@uwr.edu.pl (theory)

**ESI1. Literature data concerning the preparation of iron oxide films on silver single crystal supports (chronological order).**

| Ref. | Iron oxide | Support | Preparation procedure | Authors' description | Methods |
|---|---|---|---|---|---|
| 1 | FeO(111), Fe$_3$O$_4$(111) | Ag(111) | Repeated cycles of submonolayer iron deposition and oxidation: FeO(111) for initial 3-4 MLs, Fe$_3$O$_4$(111) for thicker films | No detailed description | LEED, XPS, XPD |
| 2 | FeO(111) (< 6 ML), Fe$_3$O$_4$(111) (> 6 ML) (also denoted as "FeO$_x$") | Ag(111) | 1-10 MLs Fe at RT; 10$^{-5}$ Torr O$_2$ at 350°C for 15 min; Cooled down in O$_2$ to 200°C; UHV annealed at 400°C for 30 min | Poorly ordered FeO(111), broadening of LEED spots and background increase, XPS indicated mixed FeO and Fe$_3$O$_4$ oxide phases or non-stoichiometric oxides | LEED, XPS, XPD |
| 2 | FeO(111) (< 10 Å), Fe$_3$O$_4$(111) | Ag(111) | Multiple cycles ≤ 0.5 ML Fe at RT; Heat to 150°C (prior to O$_2$ | Good crystalline quality FeO(111), | LEED, XPS, XPD |

| | | | introduction) 10$^{-5}$ Torr O$_2$ at 150°C for 5 min UHV annealed at 400°C for 30 min | "split" LEED beams for FeO(111), (2×2) LEED for Fe$_3$O$_4$, for thicker films XPS indicated mixed FeO and Fe$_3$O$_4$ oxide phases | |
|---|---|---|---|---|---|
| | (> 10 Å) (also denoted as "FeO$_x$") | | | | |
| 3 | FeO(001) | Ag(001) | 22 MLs $^{57}$Fe at RT in 1×10$^{-7}$ mbar O$_2$ UHV annealed at 600°C for 10 min | Stoichiometric FeO(001) with good crystalline quality, well defined LEED, only Fe$^{2+}$ in XPS, ~15% of Fe$_3$O$_4$ underneath FeO | LEED, Mössbauer |
| 4 | FeO(001), FeO(111) | Ag(001) | Not mentioned | Nearly bulk terminated FeO(001) and FeO(111) | LEED |
| 5 | FeO(111) (< 10 Å), Fe$_3$O$_4$(111) (> 10 Å), Fe$_2$O$_3$(0001) (from Fe$_3$O$_4$(111)) | Ag(111) | Multiple cycles of Fe deposition (amount not indicated) at temperatures from RT to 150°C Oxidation: 2×10$^{-6}$ mbar O$_2$ at temperatures up to 400°C (FeO(111)) 5×10$^{-6}$ mbar O$_2$ at 150°C for 15 min (Fe$_3$O$_4$(111)), UHV annealed at > 350°C 9×10$^{-6}$ mbar O$_2$ at 450°C (flash) (Fe$_3$O$_4$(111) → Fe$_2$O$_3$(0001)) | FeO(111) with lattice constant close to bulk value, well ordered epitaxial Fe$_3$O$_4$(111) and Fe$_2$O$_3$(0001) films in two domains | LEED, XRD, Raman |
| 6 | FeO(111) (< 10 Å), Fe$_3$O$_4$(111) (> 10 Å), Fe$_2$O$_3$(0001) (from Fe$_3$O$_4$(111)) | Ag(111) | Same as in Ref. 4 | This article is an extended version of Ref. 4, phase transformations between Fe$_2$O$_3$ and Fe$_3$O$_4$ possible | SXRD, LEED, Raman |
| 7 | FeO(001), FeO(111), Fe$_3$O$_4$(001) | Ag(001) | Fe deposition at RT 10$^{-5}$ mbar O$_2$ at 300°C for 1h | FeO(001) wetting the surface, Fe$_3$O$_4$(001)-like structure, quasi-hexagonal FeO(111) | LEED, STM, XPS |
| 7 | FeO(111) | Ag(001) | Fe deposition at 200-400°C in 10$^{-5}$ mbar O$_2$ | Quasi-hexagonal FeO(111) | LEED, SPA-LEED, XPS, AES, STM |
| 8 | FeO(100), Fe$_3$O$_4$ | Ag(100) | 4-80 MLs $^{57}$Fe deposition at RT (or higher) in 5.0×10$^{-8}$ mbar to 4.6×10$^{-7}$ mbar O$_2$ UHV annealed at 420-660°C for 1-7 min | Well defined FeO/Fe$_3$O$_4$ samples with different phase content ratio | XPS, Mössbauer, LEED |
| 9 | FeO(111) | Ag(001) | 0.5 ML at 570 K | Quasi-hexagonal | XPS, STM, |

| | | | in $10^{-5}$ mbar $O_2$ | FeO(111) | LEED |
|---|---|---|---|---|---|
| 10 | FeO(111), mixed FeO(111)/FeO(100), $Fe_2O_3$ | Ag(100) | Fe deposition at 25-400°C in $2×10^{-7}$ mbar, $1×10^{-6}$ mbar and $5×10^{-6}$ mbar $O_2$ | ML FeO(111) with FeO(100)-like grain boundaries, mixed FeO(111)/FeO(100) structure, hexagonal multilayer with buckled top layer – attributed to $Fe_2O_3$ | STM, LEED, XPS, NEXAFS |
| 11 | $α$-$Fe_2O_3$(0001), $Fe_3O_4$(111) | Ag(111) | Cycles of 7-9 MLs Fe deposition at RT $2×10^{-6}$ mbar $O_2$ at 720 K ($Fe_3O_4$) $2.4×10^{-5}$ mbar at 670 K ($Fe_2O_3$) | Well-ordered films, phase transformations possible | LEEM, LEED |
| 12 | FeO(111), FeO(100) (also denoted as "$FeO_x$") | Ag(100) | 0.4-2.0 MLs at 100°C in $1×10^{-5}$ mbar to $2×10^{-7}$ mbar $O_2$ UHV annealed at 400°C for 2 min | p(2×11) / c(2×12) unit cell coincidence structures, multilayer FeO | STM, LEED, XPS, NEXAFS |
| 13 | $α$-$Fe_2O_3$(0001), $γ$-$Fe_2O_3$(111), $Fe_3O_4$(111) | Ag(111) | Same as in Ref. 10, $3×10^{-5}$ mbar at < 620 K for 10 min ($γ$-$Fe_2O_3$(111)) | Well-ordered films, transformations between phases possible | LEEM, LEED, XPEEM |
| 14 | $Fe_3O_4$(111) | Ag(001) | Fe deposition at RT in $2×10^{-6}$ mbar $O_2$ UHV annealed at 675 K | $Fe_3O_4$(111) growing in two crystallographic domains | LEED, SXRD, STM |
| 14 | $Fe_3O_4$(001) | Ag(001) | Fe deposition at RT $~10^{-6}$ mbar $O_2$ at RT for 10 min $~2×10^{-7}$ mbar $O_2$ at 650 K for 30 min | $Fe_3O_4$(001) with a ($\sqrt{2}×\sqrt{2}$)R45° reconstruction | LEED, SXRD, STM |
| 14 | $Fe_3O_4$(001) | Ag(001) | Fe deposition at RT $~3×10^{-7}$ mbar $O_2$ at RT for 10 min $~3×10^{-7}$ mbar $O_2$ at 650 K for 30 min and 770 K for 30 min | Seed layer for the growth of a well-ordered (001) magnetite layer | LEED, SXRD, STM |
| 15 | FeO(111), $Fe_3O_4$(111), multilayer $FeO_x$ | Ag(111) | Fe deposition at various substrate temperatures (RT to 773 K) in $2×10^{-7}$ Torr to $1×10^{-6}$ Torr $O_2$ | FeO(111) with a (9×9) Moiré super-structure, $Fe_3O_4$(111), FeO(111)-like multilayer structure denoted as "$FeO_x$" | LEED, TPD, RAIRS |
| 16 | FeO(111) | Ag(100) | Same as in Ref. 11 | Same as in Ref. 11 | STM, LEED, RAIRS, TPD, DFT+U |
| 17 | FeO(111) | Ag(100), Ag(111) | Fe deposition at 100°C in $2×10^{-7}$ mbar $O_2$ UHV annealed at 400°C | Monolayer FeO(111) with coincidence structures same as in Ref. 15 | STM, LEED, SXRD, TPD, RAIRS |

| This work | FeO(111) | Ag(111) | 1 ML Fe at 500-600 K $1\times10^{-6}$ mbar $O_2$ at 700 K for 30 min | Moiré-free stoichiometric well-ordered FeO(111) | STM, LEED, XPS |
|---|---|---|---|---|---|

"ML" indicates "monolayers", "RT" – "room temperature", "UHV" – ultra-high vacuum. The scientific units are the same as in the original articles.

**ESI2. Detailed description of the experimental apparatus, procedures and theoretical calculations.**

The experiments were performed in an ultra-high (UHV) system consisting of three inter-connected chambers: the preparation chamber (base pressure: $5\times10^{-10}$ mbar), scanning probe microscopy chamber (base pressure: $3\times10^{-11}$ mbar) and load-lock chamber. The preparation chamber is equipped with a cold cathode sputter gun, single electron beam evaporator, e-beam heating stage, low energy electron diffraction (LEED) and x-ray photoelectron spectroscopy (XPS). The scanning probe microscopy chamber is equipped with a commercial variable-temperature scanning tunneling microscopy/atomic force microscopy (VT-STM/AFM) instrument. All STM measurements were performed in constant current mode using W tips. The Ag(111) single crystal (purity: 99.999%, polishing accuracy: < 0.1°; from MaTeck) was cleaned by repeated cycles of 1 keV and 0.6 keV argon (purity: 99.999%; Messer) $Ar^+$ ion sputtering, annealing in $O_2$ (purity: 99.999%, from Messer) under UHV and at T > 700 K. Iron (purity: 99.995%; from Alfa Aesar) was evaporated from a 2 mm rod onto a Ag(111) substrate kept at 500-600 K, with a deposition rate of 2 monolayers (MLs) per min., where 1 ML is defined as the amount of iron that would cover the surface with a closed bcc-(110) film. The deposition rates were calibrated using STM (calibration accuracy: 10%). Gwyddion[18] and WSxM[19] computer software were used for STM image processing. The oxidation of iron was performed by backfilling the preparation chamber with molecular oxygen using a leak valve. Following 15-30 min oxidation, the sample was cooled down in oxygen for several minutes. The temperature of the sample was measured using an infrared pyrometer focused on a tantalum sample holder on which the sample was mounted. The scanning tunneling spectroscopy (STS) dI/dz experiments were performed using a lock-in amplifier by applying a modulation voltage of 60 mV at a frequency of 6777 Hz to the z-piezo of the STM (which resulted in a periodic tip-sample distance change of +/- 0.1 nm). The STM bias voltage was set to 50 mV. For each measurement, the tip was first retracted from the sample's surface by 1 nm, to attenuate any tip-sample interactions, and then approached to the surface with a speed of 1 nm/sec. Each of the presented curves was averaged from 15 similar measured curves and smoothed using the locally weighted scatterplot (LOESS) method. The XPS measurements were performed using a monochromatic Al Kα (1486.6 eV) X-ray source and a semispherical electron energy analyzer operating at a pass energy of 50 eV (survey) and 20 eV (regions). The data were calibrated with respect to the Ag 3d5/2 peak (368.2 eV)[20] and fitted using CasaXPS computer software (Casa Software Ltd). A linear combination of Gauss and Lorentz functions (the so called Voigt function) and Shirley background subtraction were used for the fittings.

The performed calculations were based on the density functional theory (DFT), as implemented in the Vienna *ab initio* simulation package (VASP).[21,22,23] The electron–ion core interactions were represented by the projector-augmented wave (PAW) potentials,[24,25] with Ag $4d^{10}5s^1$, Fe $3d^74s^1$ and O $2s^22p^4$ states considered as valence states. A plane wave basis set with a kinetic energy cutoff of 400 eV was applied. The exchange-correlation energy was treated at the spin-polarized generalized gradient approximation (GGA) level, using the Perdew-Burke-Ernzerhof (PBE) functional.[26] The Brillouin zone was sampled using a Γ-centered k-point meshes. A Fermi surface broadening of 0.2 eV was applied to improve convergence of the solutions, using the second order Methfessel-Paxton method.[27] To account for the strongly correlated 3d electrons localized on the Fe ions, the Hubbard U correction was applied (GGA+U) within the rotationally invariant approach of Dudarev et al.[28], with effective parameter $U_{eff}$ = U–J = 3.0 eV, where (U,J) = (4,1) eV are the Coulomb and screened exchange parameters, respectively. The $U_{eff}$ value was previously found to provide a satisfactory description of bulk characteristics of FeO.[29] Convergence threshold for the total energy of the studied systems was set to $10^{-6}$ eV. The lattice parameter of bulk fcc Ag, *a* = 4.155 Å, calculated using 16×16×16 k-points mesh, was found to agree well with other GGA calculations performed using a similar computational method (e.g. 4.16 Å[30] or 4.14 Å[31]), and overestimated the experimental value. 4.086 Å,[32] by less than 1.7 %. A silver (111) substrate was simulated by an asymmetric slab consisting of four atomic layers of Ag with a 1×1 and 2×2 surface unit cell. In order to check the appropriateness of the size of the unit cell, additional calculations for larger supercells were also performed. The slabs were separated from their periodic images by a vacuum region of 18 Å. For surface calculations, a Γ-centered 16×16×1 k-point mesh was applied for surface 1×1 unit cell, which was appropriately reduced for larger cells. The positions of atoms in bottom two Ag layers were frozen and the atomic positions of the remaining atoms were optimized until the residual Hellman-Feynman forces on atoms were smaller than 0.01 eV/Å. The optimization of the bare surface structure yielded negligibly small expansion of the topmost interplanar separation and an 0.8% contraction of the second interlayer distance. The work function was calculated as the difference between the electrostatic potential energy in the vacuum region and the Fermi energy of the slab. A dipole correction was applied to compensate for the asymmetry of the slab and obtain the correct work function value.[33] The calculated work function of the clean Ag(111) surface, 4.49 eV, was very close to the experimental value of 4.46±02 eV,[34] and the value obtained in other calculations, 4.50 eV,[35] using similar computational method. FeO(111) monolayer was adsorbed on one side of the slab. The calculated in-plane lattice constants of 2.938 Å (PBE) and 3.032 Å (PBE+U) were used[36] for the Ag(111) and FeO(111) calculations, respectively. The system was re-optimized after the deposition of an FeO layer within the applied surface unit cell. The electron charges on the atoms were calculated using Bader analysis.[37,38] The STM images were simulated using the Tersoff-Hamann method.[39]

**ESI3. Detailed x-ray photoelectron spectroscopy data.**

| | 1 ML FeO(111)/Ag(111) – raw data | | | | | | 1 ML FeO(111)/Ag(111) - Ag 1s subtracted | | | | | | |
|---|---|---|---|---|---|---|---|---|---|---|---|---|---|
| | Fe $2p_{3/2}$ | | | Fe $2p_{1/2}$ | | | Fe $2p_{3/2}$ | | | Fe $2p_{1/2}$ | | | O 1s |
| Component | $Fe^{2+}$ | Multiplet Splitting | Satellite | $Fe^{2+}$ | Multiplet splitting | Satellite | $Fe^{2+}$ | Multiplet splitting | Satellite | $Fe^{2+}$ | Multiplet splitting | Satellite | $O^{(2-)}$ |
| Position (eV) | 710.1 | 711.9 | 714.3 | 724.3 | 727.9 | 732.3 | 710.3 | 713.2 | 717.5 | 723.9 | 727.4 | 731.2 | 529.7 |
| FWMH | 2.73 | 2.73 | 2.73 | 2.73 | 2.73 | 2.73 | 2.85 | 2.85 | 2.85 | 2.85 | 2.85 | 2.85 | 1.55 |
| Area | 565.7 | 376.4 | 298.5 | 282.8 | 188.2 | 149.3 | 672.1 | 202.8 | 79.8 | 341.8 | 115.0 | 40.0 | 608.7 |
| Concentration (%) | | | | | | | | | | | | 50.3% | 49.7% |

Both fittings confirmed that iron is in the $Fe^{2+}$ oxidation state. The presence of another $Fe^{2+}$ containing iron oxide phase – $Fe_3O_4$ – can be excluded, as it consists of a mixture of $Fe^{2+}$ and $Fe^{3+}$ ions and was reported not to exhibit an Fe $2p_{3/2}$ satellite peak.[40,41] $Fe_2O_3$, on the other hand, consists of iron in the $Fe^{3+}$ oxidation state only.

**ESI4. Geometrical details of the calculated and experimentally observed systems.**

Separations of atomic layers along the coordinate perpendicular to the (111) surface plane calculated (GGA+U) for the monolayer FeO(111) film on Ag(111), the bilayer FeO(111) on Ag(111) and a free-standing 3-monolayers-thick FeO slab, were obtained from calculations applying 1×1 surface unit cells and compared with experimentally determined values and literature values for bulk FeO. The numbering of atomic layers is referenced to the topmost Ag layer. The numbers in brackets refer to the geometry parameters resulting from structural optimization by means of experimental lattice constants of Ag and PBE+U lattice constant of FeO [square brackets]. In a free-standing 3 ML FeO slab, the layers are numbered from top (terminating O layer) to the bottom. The in-plane atomic Fe-Fe and O-O distances in FeO bulk layers are 3.04 Å, and the separation of Fe and O planes is 1.25 Å.[42]

| System | Layers | Separation (all values in Å) | |
|---|---|---|---|
| | | Calculated | Experiment |
| Monolayer FeO(111)/Ag(111) | O1-Fe1 | 0.82 (0.86) | ≈3.3 (total) (STM) |
| Monolayer FeO(111)/Ag(111) | Fe1-Ag1 | 2.35 (2.36) | |
| Bilayer FeO(111)/Ag(111) | O2-Fe2 | 0.86 (0.89) | ≈2.5 (total) (STM) |
| Bilayer FeO(111)/Ag(111) | Fe2-O1 | 1.44 (1.44) | |
| Bilayer FeO(111)/Ag(111) | O1-Fe1 | 1.29 (1.33) | ≈3.3 (total) (STM) |
| Bilayer FeO(111)/Ag(111) | Fe1-Ag1 | 2.23 (2.26) | |
| 3ML FeO slab | O1-Fe1 | 0.86 [0.79] | – |
| 3ML FeO slab | Fe1-O2 | 1.43 [1.41] | – |
| FeO(111) surface | In-plane lattice constant | PBE: 2.938 (fixed) (Exp.: 2.889) (fixed) [PBE+U: 3.032] (fixed) | 2.89 (LEED) 2.94 ± 0.08 (STM) |

The calculated and measured in-plane lattice constant values differ slightly from the value expected for bulk FeO(111), however, within the limit of error, they correspond to the surface lattice constant of Ag(111), thus indicating that the oxide layer adopts the lattice constant of the Ag support. For FeO(111) films grown on various substrates close-packed metal surfaces, an expansion or contraction of the lattice constant with respect to bulk FeO was always observed, however, accompanied by a Moiré superstructure which is not present in our system. It has to be mentioned that $Fe_3O_4$(111) or $Fe_2O_3$(0001) surfaces would show atomic periodicities of 6 Å and 5 Å, respectively.[42]

**ESI5. Calculated Bader charges and magnetic moment on atoms.**

Bader charges and magnetic moments on atoms within different layers. The charge on atom is calculated as the difference between the calculated Bader charge on atom of a given layer minus the charge on respective Fe (6.55 |e|) or O (7.45 |e|) atom in bulk FeO, and on Ag1 atom of the clean Ag(111) slab (11.02 |e|), respectively. The magnetic moment on individual Fe atoms of bulk FeO is ±3.65 $\mu_B$, and zero on O atoms. The labeling of the layers refers to that shown in Figure 3 in the main text. The entries in the column labeled as "3 ML slab" refer to the free-standing 3-MLs-thick FeO(111) slab, calculated using PBE lattice constant of Ag and PBE+U lattice constant of FeO.

| Layer | Bader charge (in \|e\|) | | | Magnetic moment ($\mu_B$) | | |
|---|---|---|---|---|---|---|
|  | 1 ML | 2 ML | 3 ML slab | 1 ML | 2 ML | 3 ML slab |
| O3 |  |  | -0.25 (-0.23) |  |  | 0.46 (0.44) |
| Fe3 |  |  | -0.43 (-0.44) |  |  | 4.16 (4.18) |
| O2 |  | -0.25 | -0.04 (-0.06) |  | 0.45 | 0.07 (0.09) |
| Fe2 |  | -0.41 | -0.02 (0.00) |  | 4.15 | -3.60 (-3.58) |
| O1 | -0.24 | -0.07 | -0.03 (-0.05) | 0.39 | 0.05 | 0.00 (0.01) |
| Fe1 | 0.04 | 0.51 | 0.76 (0.78) | 3.85 | -3.41 | 3.48 (3.48) |
| Ag1 | 0.17 | 0.19 |  | 0.00 | 0.02 |  |

**ESI6. Calculated work function values.**

| Calculated work function values (eV) | |
|---|---|
| Ag(111) (1×1) | 4.49 |
| 1ML FeO/Ag(111)-FM (1×1) | 6.56 |
| 1ML FeO/Ag(111)-FM ($\sqrt{3}\times\sqrt{3}$) | 6.57 |
| 1ML FeO/Ag(111)-AFM (same results for 1×1 and $\sqrt{3}\times\sqrt{3}$) | 5.87 |
| 2ML FeO/Ag(111)-AFM (1×1) | 6.67 |
| 2ML FeO/Ag(111)-AFM ($\sqrt{3}\times\sqrt{3}$) | 6.69 |
| 3ML FeO slab-AFM (same results for 1×1 and $\sqrt{3}\times\sqrt{3}$) | 6.95 |

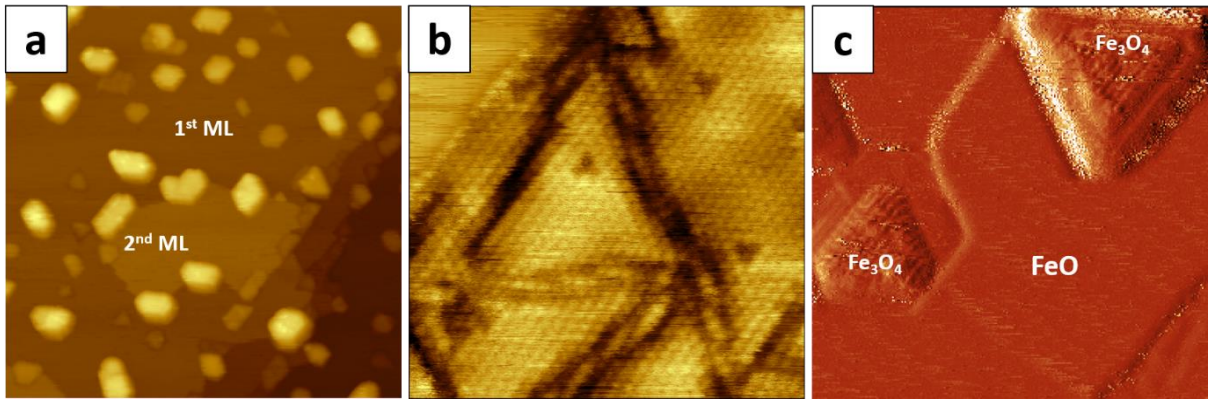

**Fig. S1** STM topography image (100×100 nm$^2$) showing the second monolayer of FeO(111) on Ag(111) (a); (b) shows STM image (10×10 nm$^2$) revealing ~3 Å periodicity and dislocation lines inside a larger defect in FeO(111) film on Ag(111); (c) presents current image (50×50 nm$^2$) of Fe$_3$O$_4$ crystals nucleating on FeO(111)/Ag(111) (V = +0.7 V, I = 0.4 nA).

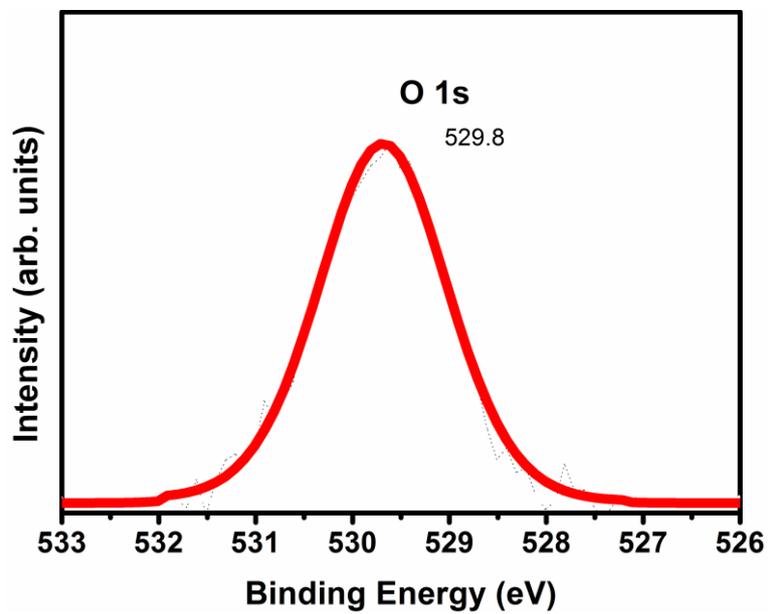

**Fig. S2** XPS O 1s spectrum obtained for a monolayer FeO(111) film on Ag(111).

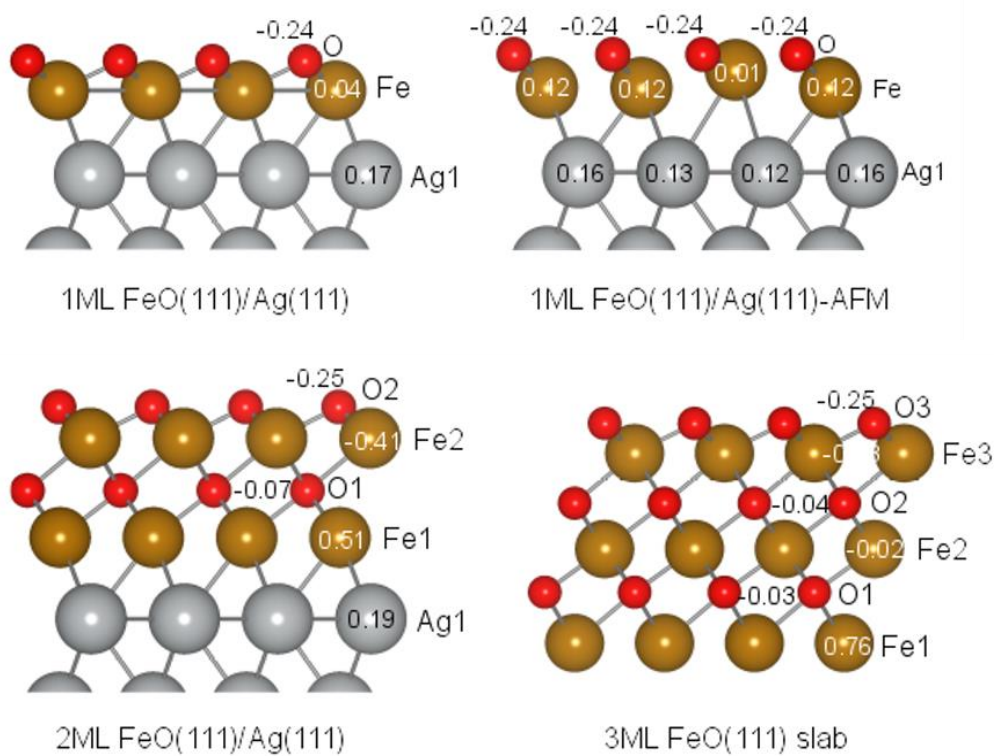

**Fig. S3** Calculated Bader charges on atoms of different FeO systems (see ESI5).

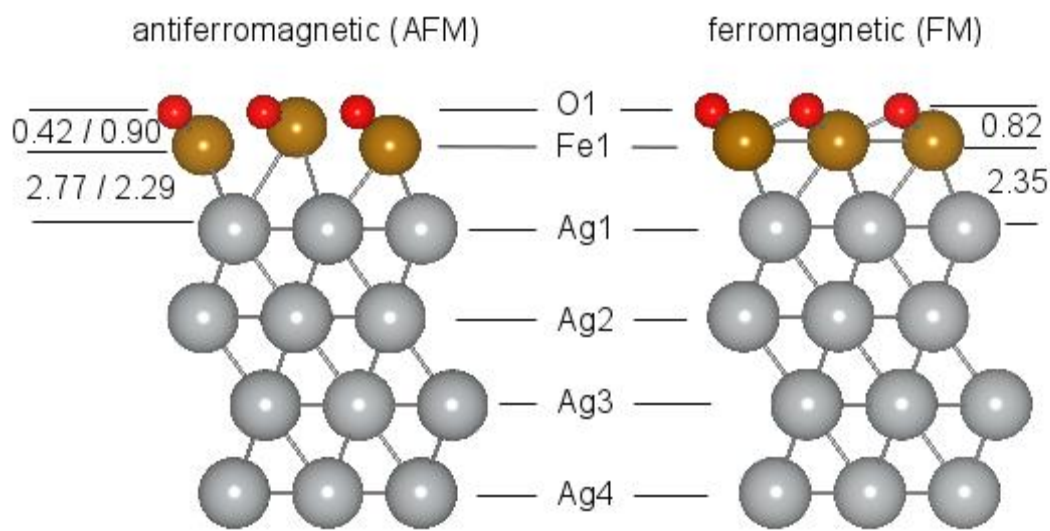

**Fig. S4** Differences in geometry of antiferromagnetic (AFM) and ferromagnetic (FM) superstructures of 1ML FeO(111) on Ag(111). The AFM alignment of magnetic moments on Fe atoms makes the Fe layer distinctly rumpled, with Fe atoms of one direction of magnetization being by about 0.5 Å more distant from the Ag(111) surface than those with opposite magnetization. All distances are given in Å.

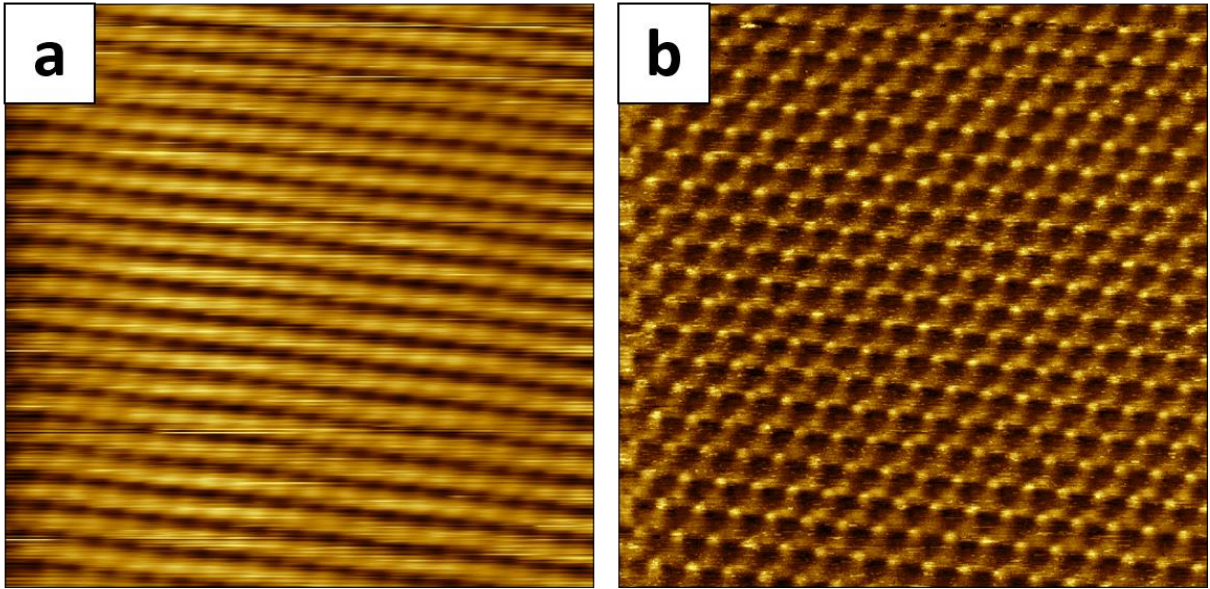

**Fig. S5** Atomic-resolution STM images (5×5 nm$^2$) of FeO(111)/Ag(111): constant-current topography (a) and current map (b) (V = +0.1 V, I = 0.4 nA).

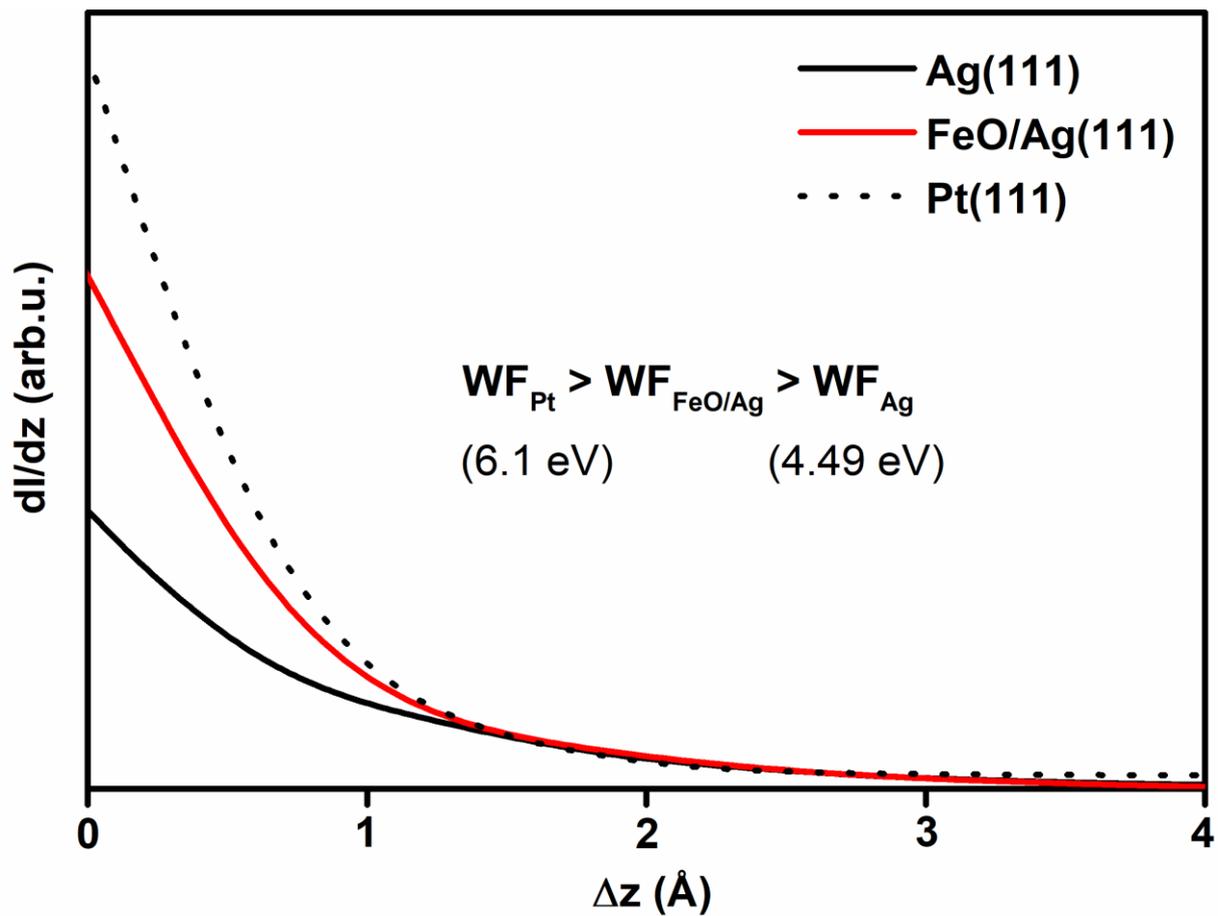

**Fig. S6** dI/dz curves obtained for FeO(111)/Ag(111), clean Ag(111) and Pt(111).

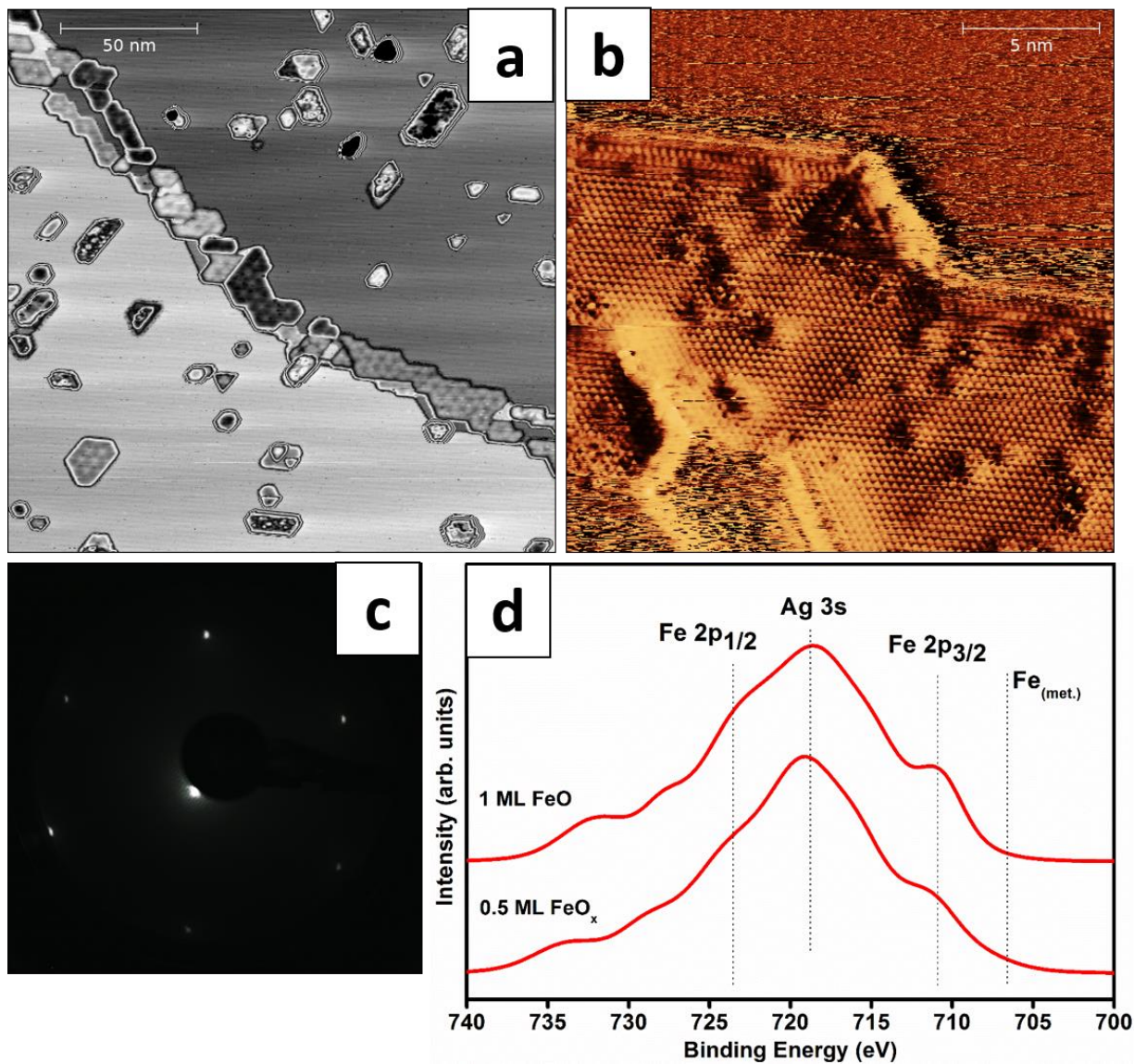

**Fig. S7** FeO$_x$ sample prepared by 0.5 ML iron deposition at room temperature onto Ag(111) and subsequent oxidation in 1×10$^{-6}$ mbar O$_2$ at T > 700 K: (a) large-scale STM image (200×200 nm$^2$, V = +0.7 A, I = 0.4 nA, presented in negative) showing ill-defined, not fully oxidized iron particles, as well as FeO$_x$ islands with a long-range superstructure; (b) atomic resolution STM image (20×20 nm$^2$, V = +0.1 V, I = 0.4 nA) exhibiting a ~3 Å periodicity on the superstructure island shown in (a); (c) shows the corresponding (1×1) LEED pattern (60 eV) and (d) the XPS Fe 2p spectrum compared with a signal from an FeO(111) film prepared by Fe deposition at 550K and post-oxidation. The similarity between FeO$_x$ and FeO(111) in LEED and XPS is contrasted by a completely different surface morphology observed in STM.